\documentclass[journal]{IEEEtran}
\ifCLASSINFOpdf
\else
\fi
\usepackage[accsupp]{axessibility}
\usepackage[dvipsnames]{xcolor}

\usepackage{url}            
\usepackage{xcolor}         

\usepackage[most]{tcolorbox}
\tcbset{size=title, frame hidden, colback=white!98!blue!98!green, colframe=white!85!blue!95!green}

\usepackage{ragged2e}

\usepackage{cite}
\definecolor{bupt-blue}{rgb}{0,0.2,0.6}
\definecolor{cvpr-blue}{rgb}{0.21,0.49,0.74}
\definecolor{dark-red}{rgb}{0.6,0,0}
\usepackage[colorlinks=true, linkcolor=dark-red, citecolor=blue]{hyperref}

\usepackage{stackengine}
\usepackage{tabularx}
\usepackage{threeparttable}
\usepackage{arydshln}

\usepackage{pifont}

\usepackage{adjustbox}
\usepackage{enumitem}

\usepackage{times}
\usepackage{microtype} 
\usepackage{graphicx} 
\usepackage{wrapfig}
\usepackage{balance}

\definecolor{p1}{HTML}{D21504}  
\definecolor{p2}{HTML}{FF00FF}  
\definecolor{p3}{HTML}{EEC600}  
\definecolor{p4}{HTML}{0FE5E1}  

\definecolor{pink}{rgb}{1 0.753 0.796}  
\definecolor{prim}{rgb}{0.965 0.914 0.945} 
\definecolor{aquaisland}{rgb}{0.659 0.855 0.863} 
\definecolor{whitelilac}{rgb}{0.945 0.914 0.965} 
\definecolor{ecruwhite}{rgb}{0.945 0.965 0.914}  
\definecolor{teal}{rgb}{0 0.502 0.502}  
\definecolor{aquaisland}{rgb}{0.6352 0.851 0.8078}   
\definecolor{sidecar}{rgb}{0.9686 0.9059 0.8078} 
\definecolor{peachcream}{rgb}{1 0.9451 0.8784} 
\definecolor{catskillwhite}{rgb}{0.9529 0.9765 0.9765} 
\definecolor{botticelli}{rgb}{0.8235 0.9098 0.9098} 
\definecolor{junglemist}{rgb}{0.6824 0.8314 0.8314}
\definecolor{neptune}{rgb}{0.5412 0.7530 0.7530}
\definecolor{azalea1}{rgb}{0.9686 0.7922 0.7883}
\definecolor{azalea2}{rgb}{0.9804 0.8588 0.8471}
\definecolor{whitelilac}{rgb}{0.945 0.914 0.965} 
\definecolor{prelude}{rgb}{0.8235 0.7059 0.8706}

\usepackage{algorithm,algpseudocode}

\usepackage{helvet}
\usepackage{courier}
\usepackage{makecell}
\usepackage{graphicx}
\usepackage{subfigure}
\usepackage{color}
\usepackage{amsfonts}
\usepackage{amsmath}
\usepackage{amssymb}
\usepackage{bm}

\usepackage{multirow}
\usepackage{multicol}
\usepackage{booktabs}
\usepackage{colortbl} 
\usepackage{caption} 
\usepackage[dvipsnames]{xcolor}
\colorlet{colorFst}{Green!25}       
\colorlet{colorSnd}{SpringGreen!45} 
\colorlet{colorTrd}{Yellow!30}      
\colorlet{colorLow}{darkgray!30}    
\colorlet{colorDeg}{Orange!30}      

\usepackage{ntheorem}

\newtheorem*{*thm}{Theorem}

\newtheorem*{*lemma}{Lemma}

\def\ie{\mbox{\textit{i.e.}, }}
\def\eg{\mbox{\textit{e.g.}, }}

\def\0{{\bf 0}}
\def\1{{\bf 1}}

\begin{document}

\title{\huge{Neural Coding Is Not Always Semantic: Toward the Standardized Coding Workflow in Semantic Communications}}

\author{Hai-Long~Qin,
	    Jincheng~Dai,~\IEEEmembership{Member,~IEEE},
		Sixian~Wang,~\IEEEmembership{Member,~IEEE},
		Xiaoqi~Qin,~\IEEEmembership{Senior Member,~IEEE},
		Shuo~Shao,~\IEEEmembership{Member,~IEEE},
        Kai~Niu,~\IEEEmembership{Member,~IEEE},
        Wenjun~Xu,~\IEEEmembership{Senior Member,~IEEE},
		Ping~Zhang,~\IEEEmembership{Fellow,~IEEE}

\thanks{This work was supported in part by the National Key Research and Development Program of China under Grant 2024YFF0509700, in part by the National Natural Science Foundation of China under Grant 62371063, Grant	62293481, Grant 62321001, Grant 92267301, in part by the Beijing Municipal Natural Science Foundation under Grant L232047, and sponsored by Beijing Nova Program. \textit{(Corresponding authors: Jincheng Dai and Ping Zhang.)}}

\thanks{Hai-Long Qin, Jincheng Dai, Sixian Wang, Xiaoqi Qin, Kai Niu, Wenjun Xu, and Ping Zhang are with Beijing University of Posts and Telecommunications.}

\thanks{Shuo Shao is with University of Shanghai for Science and Technology.}

\vspace{-1em}
}

\maketitle
\begin{abstract}
Semantic communication, leveraging advanced deep learning techniques, emerges as a new paradigm that meets the requirements of next-generation wireless networks. However, current semantic communication systems, which employ neural coding for feature extraction from raw data, have not adequately addressed the fundamental question: Is general feature extraction through deep neural networks sufficient for understanding semantic meaning within raw data in semantic communication? This article is thus motivated to clarify two critical aspects: semantic understanding and general semantic representation. This article presents a standardized definition on semantic coding, an extensive neural coding scheme for general semantic representation that clearly represents underlying data semantics based on contextual modeling. With these general semantic representations obtained, both human- and machine-centric end-to-end data transmission can be achieved through only minimal specialized modifications, such as fine-tuning and regularization. This article contributes to establishing a commonsense that semantic communication extends far beyond mere feature transmission, focusing instead on conveying compact semantic representations through context-aware coding schemes.
\end{abstract}

\section{Introduction}
\subsection{Background}
Wireless communication systems have undergone remarkable upgrades in transmitting technologies from the first generation (1G) to the fifth generation (5G). However, the existing 5G technologies are insufficient to support the demands of many emerging applications in beyond-5G (B5G) and the sixth generation (6G) networks, including industry 4.0, metaverse, brain-to-computer interaction, and digital twin, which require lower latencies and higher data rates far surpassing those of traditional 5G platforms. Meanwhile, the existing communication technologies have nearly reached the Shannon physical-layer capacity limit, prompting researchers to explore innovative solutions for the future wireless communications meeting the ambitious goals and services envisioned for B5G and 6G.

Motivated by the theoretical foundations about the three levels of communications laid by Weaver \cite{weaver-semantics} and the remarkable achievements of artificial intelligence (AI), particularly deep learning (DL), applied to solve complex engineering problems, the concept of semantic communication has emerged as a promising and groundbreaking solution, which has the potential to revolutionize traditional wireless communication systems, paving the way for the future next-generation networks.

Semantic communication distinguishes itself through the organic integration of advanced DL algorithms and models, training, coding, and performing in an end-to-end manner\cite{yang-semantic-communication-survey}. The most prominent property of semantic communication is its focus on the semantic meaning behind data sources, rather than the accurate transmission of bits and symbols. This paradigm enables traditional wireless communication systems to engage in more intelligent and context-aware interactions, where devices exchange not only raw data but also underlying semantics. Consequently, even in the presence of syntactic mismatches (\ie bit errors), semantic communication systems can somewhat avoid semantic errors.
This robustness is particularly advantageous when bandwidth is quite limited or the signal-to-noise ratio (SNR) is relatively low, as semantic communication systems may still perform well and potentially consume less resources.

\begin{figure*}[!t]
	\centering
	\includegraphics[width=\textwidth]{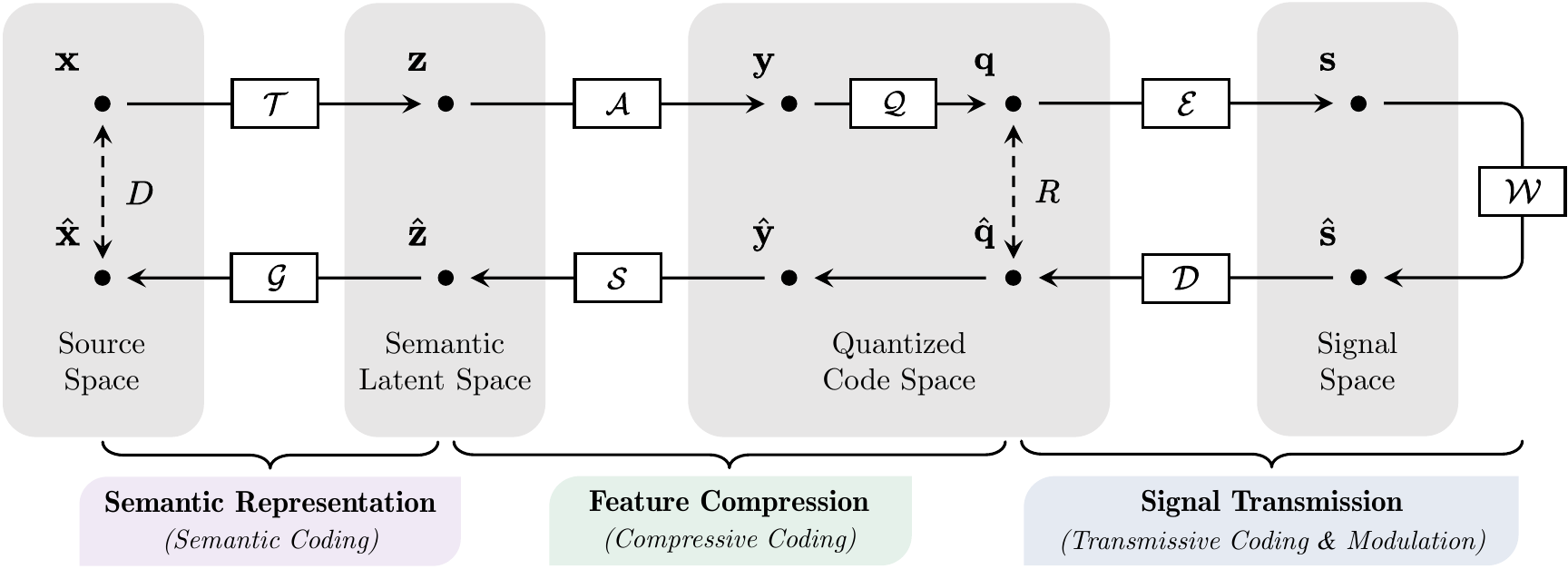}
	\caption{\textbf{Overview of the coding workflow in semantic communication systems.} 
		\emph{Notations}: Solid single arrows represent data flow, while dashed double arrows represent evaluation. $\mathcal{T}$ and $\mathcal{G}$ denote semantic tokenizer and detokenizer; $\mathcal{A}$ and $\mathcal{S}$ represent analysis and synthesis transforms; $\mathcal{E}$ and $\mathcal{D}$ indicate transmissive encoding and decoding with modulation and demodulation. $\mathcal{W}$ represents the wireless channel, $\mathcal{Q}$ denotes quantization, and $R$ and $D$ specify rate and distortion respectively. Bold lowercase letters (\eg $\mathbf{x}$) denote vectors, and vectors recovered through inverse operations are represented by bold lowercase letters with hats (\eg $\hat{\mathbf{x}}$). Unless otherwise specified, all $\mathbf{x}$ in this article represents raw data, $\mathbf{z}$ represents semantic latent representations, $\mathbf{y}$ represents neural latent features obtained through nonlinear transforms, and after quantization and entropy coding, these latent features are further compressed into bit streams $\mathbf{q}$, which are finally converted into signal sequences $\mathbf{s}$ that can be  transmitted over wireless channels through transmissive coding (\eg joint source-channel coding) and modulation. In general, semantic coding reduces dimensionality of raw data, compressive coding optimizes entropy of latent features, and transmissive coding enables symbol-to-signal conversion for transmission.}
	\label{fig:overview}
\end{figure*}

\subsection{Motivation}
Despite the prevalence of advanced AI techniques in semantic communication systems, deep neural networks (DNNs) often extract statistically sound but semantically redundant features lacking interpretable meaning. While neural coding\footnote{\textit{Neural coding} is used to denote any coding schemes built upon DNNs.} demonstrates remarkable capability in pattern differentiation, it fundamentally struggles with semantic association. The extracted features, though effective for pattern discrimination, fail to capture semantic equivalence. For example, consider a DL model trained exclusively on “corgi” images, ideally, it should recognize a “husky” image by abstracting both breeds to the higher-level category “dog” rather than treating them as entirely distinct entities. This limitation exemplifies neural coding's inability to capture genuine semantic features, stemming from inadequate concept categorization and resulting in redundant latent representations. Although neural compressive coding (\eg nonlinear transform coding \cite{balle-ntc}) achieves efficient data compression, it is not always semantic, with two critical questions unresolved:

\begin{itemize}
    \item \textbf{Question 1}: Which features are conceptually relevant or contextually meaningful for semantic communication?
    \item \textbf{Question 2}: How to extract compact general semantic representations from raw data using a unified framework?
\end{itemize}

Hence, it is quite necessary to develop a framework that goes beyond mere feature extraction and delves deeper into the semantic understanding of raw data.

In this article, we propose a novel \emph{semantic coding} framework that bridges the gap between feature extraction and semantic understanding, surpassing the limitations of previous general neural coding. One should not rashly misuse the concept of semantic coding as any coding process using neural networks. Semantic coding enables semantic communication systems to exchange and interact with the semantic meaning behind raw data through context-aware representation learning. By explicating the mechanisms through which semantics are encoded, transmitted, and decoded in wireless communication systems, semantic coding scheme facilitates a general representation of semantic meanings, striking a graceful balance between performance and efficiency.

The semantic meaning behind raw data dynamically varies in terms of different entities, making it challenging to define specific semantics with no differences for arbitrary real-world scenarios. As such, our semantic coding discussed in this article serves as a universal technical framework to explore and exploit the \emph{general semantic representations} of data sources, which represent the basic attributes and concepts behind the data. With these general semantic representations, wireless communication systems oriented towards different objects can be achieved, such as human- and machine-centric end-to-end data transmission, by integrating specialized generators (\eg generative models) or discriminators (\eg classifiers).

Our key contribution is to distinguish semantic coding from general neural coding thus achieving semantic understanding and compact representation. It can help researchers establish a commonsense: \emph{semantic communication extends far beyond mere feature transmission, focusing instead on conveying compact semantic representations through context-aware coding schemes}.

The rest of this article is outlined as follows. We begin with a preliminary overview of the comparisons between general neural coding and semantic coding, followed by the detailed introduction of how to obtain compact semantic representations. Subsequently, we explore the applications of the proposed semantic coding in both human- and machine-centric semantic communications. At the end of this article, open issues are discussed, followed by concluding remarks.

\begin{figure*}[!t]
	\centering
	\includegraphics[width=\textwidth]{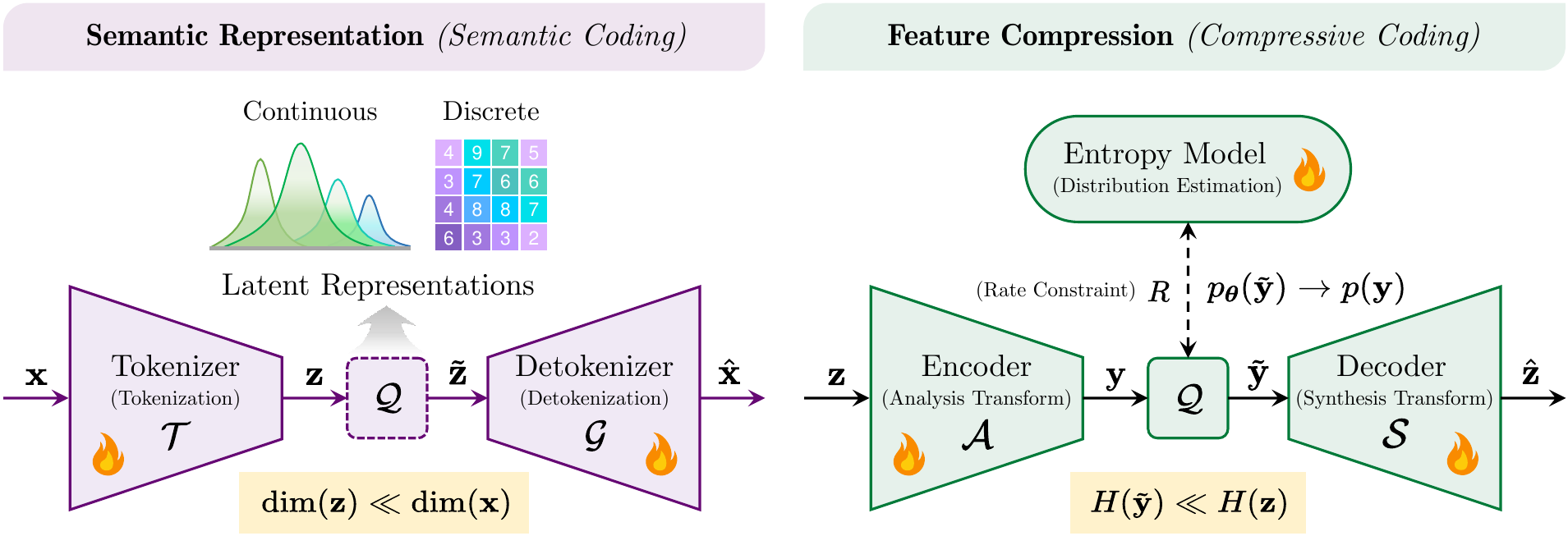}
	\caption{\textbf{Semantic coding for compact representation (towards dimensionality reduction) vs. Compressive coding for feature compression (towards entropy optimization).} \emph{Notations}: $\mathrm{dim}(\cdot)$ specify data dimensionality, and $H(\cdot)$ define information entropy. Bold lowercase letters with tildes denote quantized vectors (\eg $\tilde{\mathbf{z}}$). The dashed box represents optional operations. Compressive coding implements data compression through three components: (1) an encoder performing analysis transform $\mathcal{A}$ to convert data into a compression-friendly form, (2) a decoder executing synthesis transform $\mathcal{S}$ as the inverse operation, and (3) an entropy model predicting the probability distribution of quantized latent features in both encoding and decoding modules ($p_{\boldsymbol{\theta}}(\tilde{\mathbf{y}}) \to p(\mathbf{y})$ with model parameter $\boldsymbol{\theta}$, achieving optimized entropy $H(\tilde{\mathbf{y}}) \ll H(\mathbf{z})$). These modules, typically implemented with DNNs, are trained jointly end-to-end. For high-dimensional real-world data with unknown manifolds, entropy coding for feature compression becomes challenging. Semantic coding addresses this by serving as a pre-processing module for dimensionality reduction, obtaining low-dimensional latent representations ($\mathrm{dim}(\mathbf{z}) \ll \mathrm{dim}(\mathbf{x})$, either continuous or discrete, whose distributions are easier to estimate) before compressive coding. The tokenizer $\mathcal{T}$ extracts general semantic representations independent of individual preferences, while the detokenizer $\mathcal{G}$ adapts to specific tasks.
	}
	\label{fig:bg}
\end{figure*}
\section{From Neural Coding to Semantic Coding}

\subsection{Rethinking General Neural Coding}

As a general neural coding approach, compressive coding achieves data compression by mapping data sources to alphabet codes (typically binary streams), as illustrated in Fig. \ref{fig:overview}. This approach enables effective data recovery within acceptable distortion bounds (lossy coding) and has recently emerged as the dominant paradigm for source compression in semantic communication. In essence, it optimizes the entropy of quantized latent features through joint training of an auto-encoder and a learnable entropy model that predicts input data distributions (see Fig. \ref{fig:bg}, right). However, raw data often resides in high-dimensional manifolds with unknown and complex distributions, making effective entropy coding challenging.

While extracted latent features by such general neural coding can partially capture statistical properties of data sources, they often fall short in thoroughly exploring or effectively representing the underlying semantic meaning behind raw data, retaining substantial dimensional redundancy that complicates probability distribution estimation. From a perspective of neural network designing, existing limitations of neural coding can be primarily attributed to:

\begin{itemize}
	\item \textbf{Limited generalization capability}: General neural coding exhibits poor performance when processing out-of-distribution (OOD) data due to inductive biases.
	\item \textbf{Lack of adaptive semantic feature extraction}: General neural coding fails to dynamically identify and prioritize semantically relevant or significant features.
	\item \textbf{Insufficient contextual relationship modeling}: General neural coding often struggles to effectively capture global context and long-range semantic dependencies.
\end{itemize}

Recent advances in representation learning suggest contextual modeling as a promising solution to overcome the limitations of general neural coding \cite{dosovitskiy-vit, bolya-tome, yu-titok}. Contextual modeling projects raw data into a low-dimensional subspace within its context, enabling pre-trained DL models to extract meaningful patterns and create more interpretable representations that can adapt to new downstream tasks even without fine-tuning.

Contextual modeling is based on tokenization, which is the process of decomposing structured or unstructured data into discrete, machine-processable units (\ie tokens). Token is the \emph{basic learnable element} for data representation (not just a feature), capturing semantic meaning at variable granularities while enabling efficient contextual modeling.

Unlike naive binary digit representations ($0$/$1$ bits), token representation achieves higher information density by functioning as aggregated bit sets and forming the fundamental building blocks of context. These contexts of raw data comprise general semantics, including both structural attributes and conceptual elements. However, conventional DL models employed in neural coding are primarily designed for specific tasks based on their training data, which often exhibit limited semantic modeling capabilities due to their high dependency on training input-output pairs and lack of context-aware representation learning. Therefore, this article introduces the standardized procedure of semantic coding, an extensive neural coding scheme for general semantic representation leveraging contextual modeling in semantic communication systems.

\subsection{Steps to Attain Semantic Coding}

\begin{figure*}[!t]
	\centering
	\includegraphics[width=1.0\textwidth]{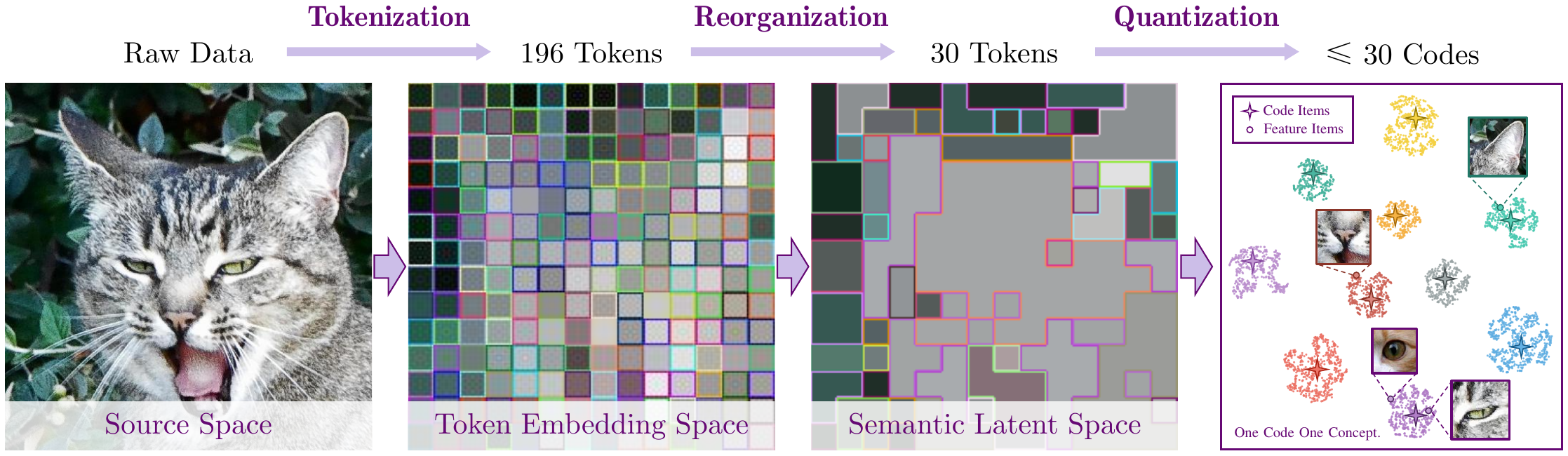}
	\caption{\textbf{Illustration of extracting general semantic representations.} Raw data in \emph{source space} is transformed into tokens via tokenization, and these obtained tokens in \emph{token embedding space} are then reorganized into compact semantic tokens constituting the so-called \emph{semantic latent space} through similarity-based consolidation. In this process, similar tokens are aggregated together, indicating high semantic relevance. Optionally, scalar or vector quantization can be applied to represent semantic tokens as latent codes retrieved from a codebook. Furthermore, synonymous concepts could be further aggregated within conceptual contexts for more generalized semantic understanding. This means sub-concepts can be further reorganized in the semantic latent space by aggregating similar ones through a process analogous to token reorganization. Subsequent quantization would yield sparse codes that reduce semantic overlap and ambiguity, approaching ``one code one concept''.}
	\label{fig:semcod}
\end{figure*}

Semantic coding \emph{reshapes the topological space of data manifolds and transforms raw data into more manageable latent representations} while reducing dimensionality, as illustrated in Fig. \ref{fig:bg}. This coding scheme serves as a pre-processing module that enables entropy models to more accurately predict the distributions of low-dimensional latent features during compression, thereby reducing bandwidth requirements in communication systems. To put it simply, semantic coding develops neural coding by incorporating contextual modeling principles through specialized operations, including \emph{tokenization}, \emph{reorganization}, and \emph{quantization} (optional), which together empower the transition from feature-level to semantic-level representation, as is illustrated in Fig. \ref{fig:semcod}.

\subsubsection{Tokenization Makes Contexts Manageable}
The latent features extracted by neural coding may be biased or unbalanced, lacking unified representation that captures long-range dependencies, which can lead to poor contextual modeling. As such, semantic coding utilizes tokenization to achieve a more holistic contextual representation of the prior distributions (\ie real data distributions), thereby mitigating the misaligned inductive biases with human cognition inherent in most DNNs employed by general neural coding. This process introduces a context-aware regularization into the compression process, directing neural networks to focus on major categorical aspects of data sources rather than overfitting to marginal or irrelevant aspects through dynamic contextual modeling.

Hence, tokenization makes contexts manageable and enhances the default inductive biases of DNNs, thus ensuring the semantic coding framework's proficiency in high-level conceptual abstraction.

\subsubsection{Attention Makes Tokenizers Flexible}
General neural coding's limitation in adaptive feature extraction often results in inaccurate and inefficient categorical prediction for complex data. Semantic coding addresses this by incorporating attention mechanisms (a necessary condition, \eg multi-head self-attention) into tokenization, enabling DL models to focus on regions with higher confidence (indicated by larger similarity scores or relation weights).

The attention mechanism enhances contextual comprehension by enabling models to dynamically weight different elements of the input token sequence based on their relevance. Rooted in psychological principles of selective concentration, attention represents the strategic allocation of limited cognitive resources to process contextual information. This fundamental property allows semantic coding to effectively extract and analyze the salient semantic meanings from raw data.

\subsubsection{Reorganization Makes Representation Compressible}
Despite this, attention-aided tokenization is still insufficient for semantic communication systems that require compact data representation for low-bandwidth transmission, \eg Transformer-based neural coding schemes, while identifying contextually relevant tokens through self-attention, does not constitute semantic coding as it lacks mechanisms for \emph{relevant token reorganization} (\eg retrieval and merging) for compact representation. Therefore, variable-length token representation strategies that adaptively consolidate tokens based on similarity should be considered.

Most current tokenizers lack such flexible reorganization capability, \eg the widely-adopted visual tokenizer in ViT \cite{dosovitskiy-vit}. The representation length in these tokenizers (\eg the number of tokens) remains fixed for all images, determined by the human-engineered patch partition, thus precluding adaptive representation or compression and resulting in suboptimal efficiency. Besides, while increasing the number of latent tokens improves image reconstruction quality, the gains diminish beyond $128$ tokens. Intriguingly, just $32$ tokens can achieve satisfactory reconstruction performance\cite{yu-titok}.

Semantic coding addresses these problems by computing token similarity in an embedding space, where proximate token embeddings maintain high contextual relevance and typically represent similar meanings or concepts. These semantically related tokens are then consolidated into a single representative token (\eg the average candidate token among neighboring tokens) for compact semantic representation. Efficient and lightweight search algorithms, such as bipartite soft matching \cite{bolya-tome}, k-nearest neighbors and locality sensitive hashing, can be employed for contextual token reorganization\footnote{Note that while this article focuses on search algorithms (clustering) for token consolidation with consistent semantic concepts, alternative approaches (\eg contrastive learning,  distilling, pooling, and latent space alignment with foundation models) can achieve similar outcomes through similarity matching.}. Given the versatility of these search algorithms, semantic coding can be implemented on pre-trained attention-aided tokenizers, enhancing their inductive biases while improving efficiency and throughput with minimal accuracy drop. It is hypothesized that token reorganization works due to its concise contextual representation in concept categorization, enabling effective attention and utilization of semantic-rich information.

For example, consider the vanilla ViT-B model with the patch size of $16\times16$ processing a $224\times224$ input image. While general neural coding requires $196$ tokens (or even more) to represent the complete image content, semantic coding can reduce this to merely $5\%$ to $15\%$ of the original amount (approximately $10$ to $30$ tokens), \emph{with a few tokens sufficient to represent one semantic concept in terms of source compression}. Consequently, the significant tokens for general semantic representation are those candidates representing numerous relative neighbors in context (assuming each semantic concept of raw data holds equal contribution), which should be prioritized for protection during wireless transmission. By the way, while token pruning offers an alternative approach for compact semantic representation, it may compromise error detection and correction capabilities.

\subsubsection{Quantization Makes Semantics Discernible}
With compact semantic tokens, quantization can be applied to generate sparse latent codes, where a single latent code can even represent one semantic concept. Quantization converts continuous values into discrete levels suitable for computer processing, and it generally takes two forms \cite{dai-deep}: scalar quantization with infinite codebook size, which rounds each latent representation to its nearest integer value, and vector quantization with finite codebook size, which matches latent representations to their closest vectors in a learned codebook. Entropy coding techniques (\eg arithmetic coding) typically follow quantization to further reduce bit rates. While quantization enables more compact semantic representations, it introduces quantization errors and increased reconstruction complexity. Therefore, one should utilize quantization optionally based on specific scenarios.

To conclude, the extraction of general semantic representations, illustrated in Fig. \ref{fig:semcod}, comprises three fundamental steps: tokenization, reorganization, and quantization. Initially, raw data is processed by the attention-aided DL model and transformed into embedded token vectors through tokenization. These tokens then undergo reorganization into compact semantic tokens through similarity computation, where the semantically consistent are identified and merged. Finally, an optional quantization step can be applied to the compact semantic tokens to generate sparse discrete latent codes, enabling further redundancy reduction and enhanced compatibility with auto-regressive (AR) based generators for content recovery\cite{yu-titok}. In practical deployments, semantic coding can be easily implemented by modifying off-the-shelf attention modules in pre-trained Transformers, supporting both task-oriented fine-tuning and model retraining.
\section{Semantic Coding towards Human-Centric Semantic Communication}  \label{sec:human-centric}

\begin{table*}[t]
	\centering
	\caption{An overview of commonly employed metrics for visual quality assessment \cite{zhai-vqa}.}\label{tab:metric}
	\renewcommand{\arraystretch}{1.5}
	\begin{tabular}{>{\noindent\justifying\arraybackslash}p{1.5cm}|>{\noindent\justifying\arraybackslash}p{1.5cm}|>{\noindent\justifying\arraybackslash}p{6.5cm}|>{\noindent\justifying\arraybackslash}p{6.5cm}}
		\hline
		\hline
		\cellcolor{whitelilac}{\textbf{Category}} & \cellcolor{prim}{\textbf{Metric}} & \cellcolor{catskillwhite}{\textbf{Introduction}} & \cellcolor{ecruwhite}{\textbf{Limitation}} \\
		\hline
		\multirow{7}{*}{FR metrics}  
		& \multirow{1.8}{*}{PSNR $\uparrow$} & Quantifies image quality by computing the logarithmic ratio of peak signal power to mean squared error. & \multirow{1.8}{*}{Sensitive to perceptual quality variations.} \\
		\cline{2-4}
		& \multirow{1.8}{*}{SSIM $\uparrow$} & Measures structural similarity between images based on luminance, contrast, and structure. & \multirow{1.8}{*}{Limited sensitivity to color or texture differences.} \\
		\cline{2-4}
		& \multirow{1.8}{*}{LPIPS $\downarrow$} & Evaluates perceptual similarity using learned deep features. & \multirow{1.8}{*}{Depends on the quality of the pre-trained network.} \\
		\cline{2-4}
		& \multirow{1.8}{*}{DISTS $\downarrow$} & Combines deep features and traditional statistics for perceptual quality assessment. & \multirow{1.8}{*}{Computationally intensive.} \\
		\hline
		\multirow{3.5}{*}{RR metrics} 
		& \multirow{1.8}{*}{RRED $\downarrow$} & Estimates quality using partial reference information and statistical models. & \multirow{1.8}{*}{Requires specific feature design.} \\
		\cline{2-4}
		& \multirow{1.8}{*}{RR-SSIM $\uparrow$} & A reduced-reference variant of SSIM based on limited structural information. & \multirow{1.8}{*}{Less effective with significant reference information loss.} \\
		\hline
		\multirow{5}{*}{NR metrics} 
		& \multirow{1.8}{*}{NIQE $\downarrow$} & Computes quality based on natural scene statistics without a reference image. & May misinterpret distortions not aligned with natural scene assumptions. \\
		\cline{2-4}
		& \multirow{1.8}{*}{FID $\downarrow$} & Measures similarity between feature distributions of real and generated images. & \multirow{1.8}{*}{Sensitive to feature extractor selection.} \\
		\cline{2-4}
		& \multirow{1.8}{*}{KID $\downarrow$} & A kernel-based alternative to FID with unbiased estimation. & \multirow{1.8}{*}{Computationally slower for large datasets.} \\
		\hline
		\hline
	\end{tabular}
\end{table*}

\subsection{Consistency-Realism Trade-off}

Immersive services, including virtual reality and volumetric video, are one of the most promising human-centric applications in next-generation wireless networks. However, current 5G technologies struggle to meet the demands of these emerging immersive applications. For example, transmitting volumetric data, particularly high-resolution and 360° videos, can require data rates reaching terabits per second (Tbps), while delivering volumetric data streams demands end-to-end latencies as low as a few milliseconds (ms) to ensure satisfactory user experience. This challenge has driven a paradigm shift from traditional wireless communication based on separate source and channel coding (SSCC) to end-to-end semantic communication leveraging joint source and channel coding (JSCC) for high-dimensional visual data compression and real-time communications (RTC) under dynamic channel conditions.

The theoretical foundations of lossy data compression stem from Shannon's rate-distortion theory, which illuminates the trade-off between bit rate for data representation and the resultant distortion in recovered data. However, minimizing distortion alone given a fixed bit rate does not guarantee good perceptual quality in reconstruction. Conversely, incorporating generative adversarial losses can enhance perceptual realism, albeit often at the expense of increased distortion\cite{blau-rdp}.

Specifically for semantic coding, such \emph{consistency-realism trade-off}\footnote{This article examines the consistency-realism trade-off specifically in visual data compression. Consistency (\ie faithfulness) measures the content consistency between the original and recovered data at receiver side. Realism (\ie naturalness) refers to the perceptual quality of recovered data.} requires flexible control over the number and dimensionality of tokens\cite{yao-rec&gen}. An excessive number of embedded tokens introduces redundancy, actually reverting to neural coding, which undermines perceptual realism and increases bandwidth overhead. On the other hand, fewer or lower-dimensional tokens do not necessarily yield a more compact and meaningful representation, as they may lack sufficient semantic richness and become difficult to decode accurately, resulting in reduced consistency. A proper choice of token quantity or dimensionality can be achieved through the \emph{reorganization} introduced earlier.

\subsection{Visual Quality Assessment}
Consistency-realism trade-off also necessitates appropriate metrics for visual quality assessment, which can be categorized into three types:

\begin{itemize}
	\item \textbf{Full-reference (FR) metrics}: These metrics focus on evaluating the consistency by comparing recovered data against a ground-truth reference, quantifying the content discrepancy between the two.
	\item \textbf{Reduced-reference (RR) metrics}: These metrics evaluate quality by comparing recovered data with partial ground-truth information, performing assessment with limited reference access.
	\item \textbf{No-reference (NR) metrics}: These metrics assess the quality of recovered data directly, evaluating its naturalness, validity, and realism without requiring a reference.
\end{itemize}

Beyond conventional full-reference metrics like peak signal-to-noise ratio (PSNR), this article also examines other commonly employed metrics for visual quality assessment evaluating consistency-realism trade-off, as detailed in Tab. \ref{tab:metric}.

\subsection{Fidelity-Driven Human Semantic Communication}
\begin{figure*}[!t]
	\centering
	\includegraphics[width=0.9\textwidth]{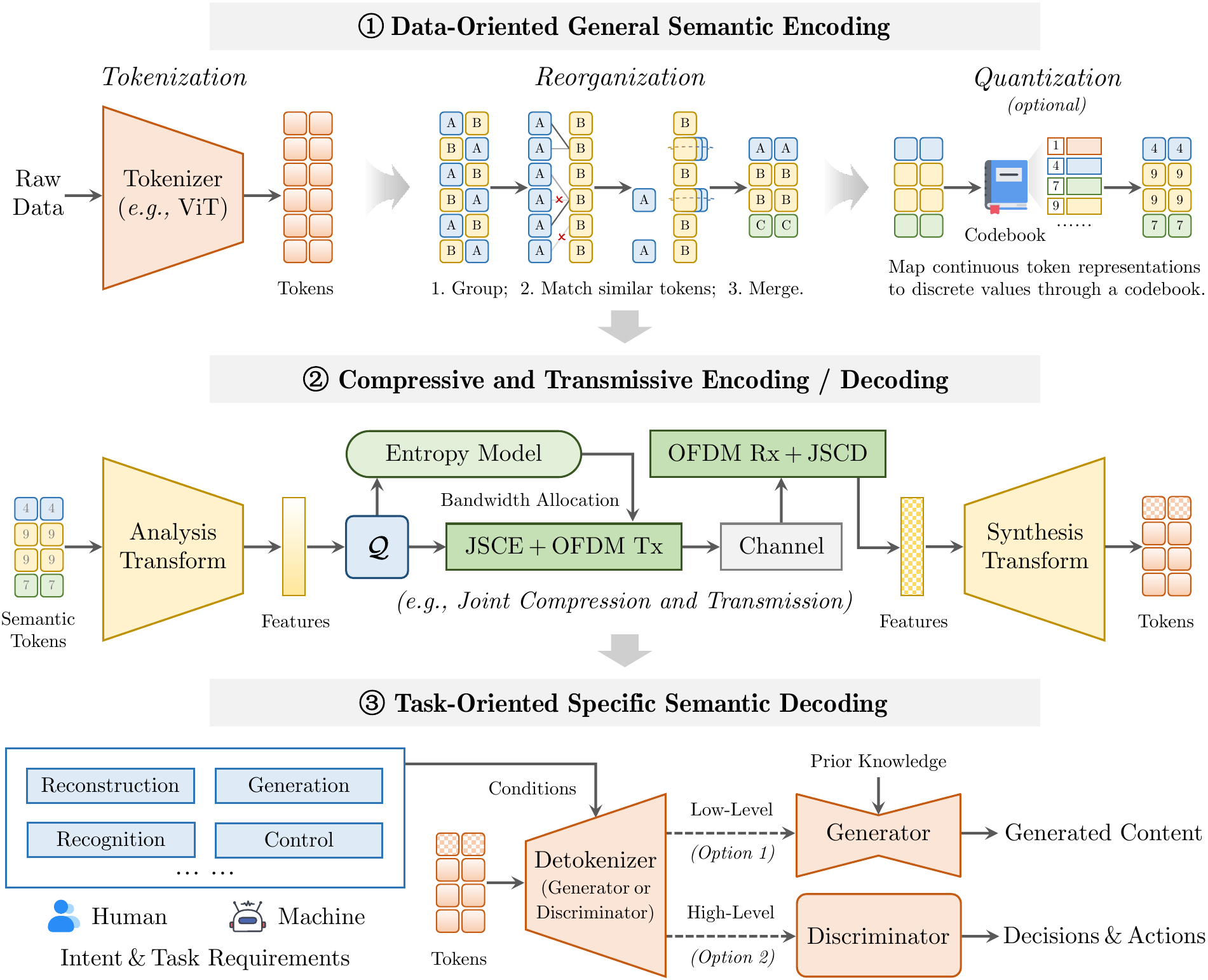}
	\caption{\textbf{The technical workflow of semantic coding enabled semantic communication.} \textit{Notations:} $\mathrm{JSCE}$ and $\mathrm{JSCD}$ denote joint source-channel encoding and decoding respectively. They constitute joint source-channel coding as an implementation of transmissive coding for real-time high-dimensional data transmission. Dashed single arrows represent optional generator or discriminator based on task levels. General semantic representations are extracted through semantic encoding via three steps: tokenization, reorganization (for compact semantic tokens, implemented following \cite{bolya-tome} with deployment ease and computational efficiency), and quantization with the optionality depending on the properties of utilized semantic detokenizers. To be specific, token reorganization is achieved through three steps: (1) group all tokens into two sets $\mathrm{A}$ and $\mathrm{B}$ of roughly equal size, (2) draw edges from each token in $\mathrm{A}$ to match its most similar token in $\mathrm{B}$ using self-attention and retain only the $r$ most similar edges (we set $r=8$), and (3) merge connected tokens to form new token set $\mathrm{C}$ with consistent semantic meaning. Subsequently, compressive and transmissive coding (\eg nonlinear transform source-channel coding\cite{dai-ntscc}) are employed for end-to-end wireless transmission. At receiver side, semantic decoding is tailored to specific human intents or machine tasks. For low-level tasks focusing on detail reconstruction, generators (\eg generative models) are employed, while for high-level tasks emphasizing recognition and decision-making, discriminators (\eg learned classifiers) are typically utilized. The intended content is recovered from general semantic representations conditioned on guidance from specialized inputs (\eg user prompts as prior knowledge).
	}
	\label{fig:architecture}
\end{figure*}

Human-centric semantic communication systems prioritize user experience by emphasizing both content faithfulness and perceptual satisfaction, \ie pursuing high fidelity, focusing on extracting semantic meaning from raw data through DNNs trained on large datasets. With appropriate visual quality assessment metrics evaluating the consistency-realism trade-off, a promising technical roadmap for end-to-end human-centric semantic communication systems based on semantic coding becomes feasible, as shown in Fig. \ref{fig:architecture}.

In this roadmap, semantic coding compresses raw data to a semantic level, followed by compressive and transmissive coding (\eg nonlinear transform source-channel coding\cite{dai-ntscc}) for efficient end-to-end wireless transmission (with modulation converting data into radio waves). At receiver side, a generator is employed to facilitate user-customized data recovery. This generator, implemented through generative modeling, can be either discrete with quantization or continuous without quantization. For example, the generator may utilize a continuous deep generative model, such as the diffusion model\cite{rombach-ldm}, which demonstrates robust and intended content generation through controllable stochastic posterior sampling. In continuous implementations, semantic coding bypasses quantization, with latent representations directly encoded, transmitted, and decoded. This contrasts with discrete implementations, where token embeddings are clustered into the nearest latent codes (\ie codewords within a codebook) as compact sparse representations at the expense of reconstruction accuracy.

Conditions on deep generative modeling play a crucial role in achieving desired content generation while maintaining an optimal consistency-realism trade-off in recovered data. Beyond user-input conditions like ``prompts'', source knowledge base\cite{ren-kb}, which provides strong prior knowledge aligned with human intent, serves as an effective guideline for precise semantic interpretation of general semantic representations, enabling user-intent-aligned data recovery at receiver side.

\subsection{A Case Study}

This section presents an empirical exploration validating the advantages of end-to-end human-centric semantic communication utilizing semantic coding. Consider a point-to-point communication system where a transmitter encodes and transmits RGB images, while a receiver performs image decoding and recovery. The system operates over an additive white Gaussian noise (AWGN) channel (SNR = $1\,$dB) simulated across multiple rate levels ($[0.0208, 0.0417, 0.0625, 0.0833, 0.125]$). The rate is measured with channel bandwidth ratio\footnote{Channel bandwidth ratio (CBR) is defined as the ratio between the transmitted signal dimension $k$ and the original signal dimension $m$, \ie $\text{CBR} = k/m$.} (CBR), \ie the channel utilization per pixel (cpp). All transmitted images are compressed into either $30$ or $10$ tokens.

\begin{figure*}[!t]
	\centering
	\includegraphics[width=\textwidth]{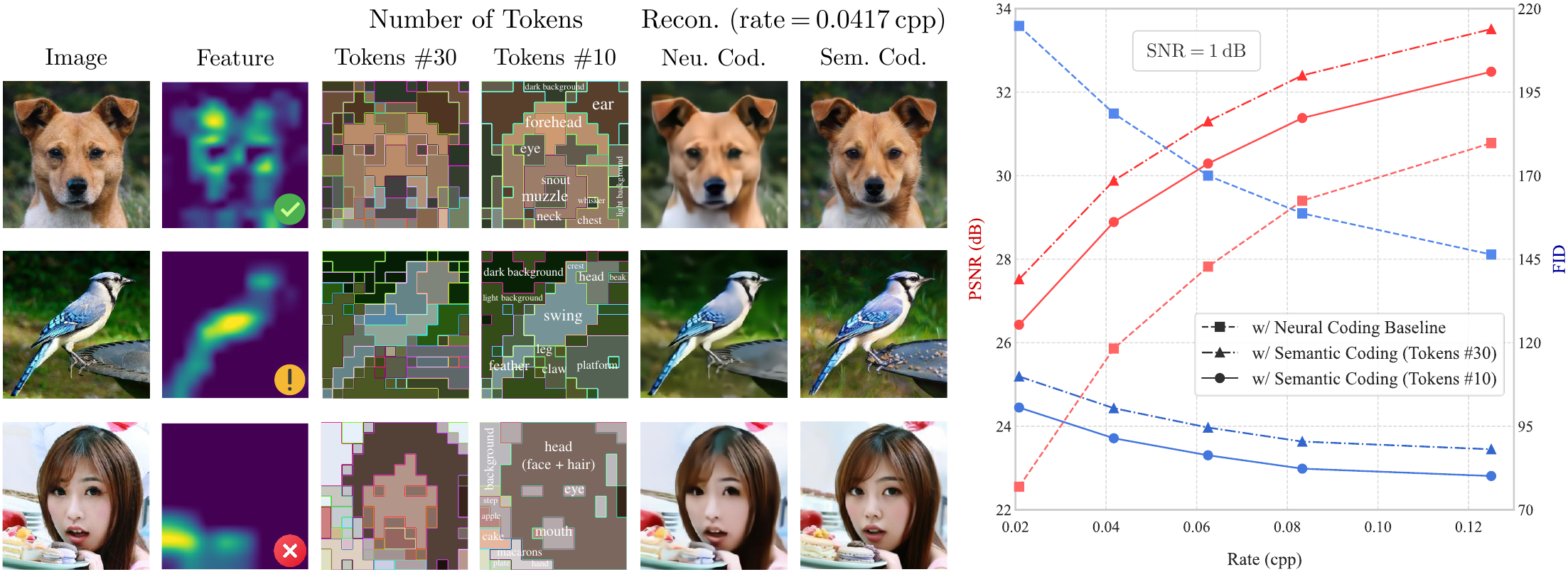}
	\caption{\textbf{Results for end-to-end wireless image transmission with semantic coding.} “$\mathrm{Recon.}$”, “$\mathrm{Neu.\ Cod.}$”, “$\mathrm{Sem.\ Cod.}$” and “w/” denote “Reconstructions”, “Neural Coding”, “Semantic Coding” and “with” respectively. A higher PSNR or lower FID score indicates better performance. Features are captured directly from neural coding baseline. The reconstructed results show semantic coding achieves realistic images while maintaining acceptable consistency. Zoom in for details.}
	\label{fig:result}
\end{figure*}

To achieve this semantic compression, we implement a bipartite soft matching module for token reorganization (matching and merging $8$ tokens per layer) between the attention and MLP branches of each Transformer block in off-the-shelf ViT-B/16 model, which is a vanilla ViT \cite{dosovitskiy-vit} with base architecture that processes images divided into $16 \times 16$ pixel patches. This semantic coding implementation requires no retraining, fine-tuning, or gradient tricks, though these optimizations remain available for enhanced performance-efficiency trade-offs. The overall JSCC framework for wireless transmission is trained end-to-end on large-scale image datasets. At receiver side, a Stable Diffusion \cite{rombach-ldm} generator performs content recovery.

For comparative analysis, we implement a well-established neural coding baseline based on \cite{bourtsoulatze-deepjscc}. While other visual quality assessment metrics mentioned, both schemes are evaluated using PSNR and FID for simplicity. As illustrated in Fig. \ref{fig:result}, the proposed semantic coding based system performs better across all CBR regimes, with a particularly significant performance gap over neural coding at low rate levels. Visual comparisons of reconstructed images at rate equaling $0.0417$ cpp reveal that semantic coding produces notably more realistic images through the generator while maintaining acceptable consistency. Furthermore, while general neural coding may extract biased or ambiguous features, semantic coding ensures clear concept categorization underpinning semantic meaning behind raw data through compact token representations. Notably, semantic coding achieves an inference FP16 throughput of $524$ images per second with a Top-1 accuracy of $74.06\%$, representing only a $5\%$ accuracy reduction while nearly $1.5 \times$ throughput compared to the vanilla ViT-B/16, let alone neural coding.

\section{Semantic Coding towards Machine-Centric Semantic Communication}  \label{sec:machine-centric}

\subsection{Efficiency-Effectiveness Trade-off}
Task-oriented semantic communication leverages an auto-encoder style transceiver, treating the wireless channel as a non-trainable neural network bottleneck layer to optimize scarce transmission resources, including bandwidth and energy. This article explores task-oriented machine-centric semantic communication systems implemented in an end-to-end manner. These systems prioritize effective task completion, including reconstruction, generation, recognition, decision (control) and so on, where only task-relevant, important, and useful information is recovered from general semantic representations at receiver side.

For machine terminals with constrained computing and storage capabilities, \eg unmanned aerial vehicles and intelligent robots, efficient data compression becomes imperative for bandwidth reduction and energy conservation. Semantic coding emerges as a promising solution, achieving competitive performance with $\leqslant 30$ tokens representing semantic meaning of a $224 \times 224$ RGB image. Additionally, system robustness and reliability are crucial for combating semantic noise, \eg adversarial perturbations, which can cause misalignment between intended and received semantic representations, leading to task failure\cite{zhang-unified}. While protection operations for resilient transmission are necessary to ensure effective task-oriented machine-centric semantic communication, they may introduce additional overhead, necessitating carefully balancing the \emph{efficiency-effectiveness trade-off}.

Furthermore, existing neural coding based task-oriented semantic communication systems require training from scratch to both adapt to changed downstream tasks and maintain compatibility across different DL models for multimodal tasks. The contextual modeling based semantic coding scheme offers promise in addressing these challenges through unified, general, compact, and meaningful semantic representations.

\subsection{Task-Driven Machine Semantic Communication}
A methodology similar to Section \ref{sec:human-centric} is applied to machine-centric semantic communication systems (see Fig. \ref{fig:architecture}), where generators or discriminators facilitate task-specific data recovery. Different downstream tasks require varying levels of recovery granularity, \eg data reconstruction demands high-fidelity restoration with minimal distortion, while recognition tasks merely require correct class label classification.

Accordingly, task-specific post-processing modules can be implemented: generators (\eg auto-regressive models) for low-level detail generation, or discriminators (\eg learning-based classifiers) for high-level category discrimination. Machine-centric semantic communication systems typically emphasize high-level tasks such as recognition and control, necessitating discriminators for decision-making and action scheduling.

Consider, for example, a static ground base station transmitting control and command (C\&C) signals (derived from general semantic representations) to an unmanned aerial vehicle through a wireless channel with line-of-sight propagation and Rayleigh fading. A deep reinforcement learning based discriminator at receiver evaluates signal importance based on C\&C parameter similarity and information age, optimizing transmission decisions to balance resource efficiency and control reliability.

Given the computational and storage constraints of machine terminals, quantization is frequently employed for compact semantic representation, reducing transmission bandwidth and machine energy consumption in end-to-end systems. These quantized semantic coding based systems can effectively combat naturally occurring semantic noise\cite{zhang-unified}, leveraging the sparsity of latent codes to abstract semantic concepts and withstand large-scale perturbations.

The robustness of quantized semantic coding becomes particularly evident when combined with generators under extremely poor channel conditions, such as in underground transportation hubs with high-density pedestrian flows. When lightweight generators are deployed on machine terminals, they can efficiently recover data from quantized sparse latent codes and generate complete, task-relevant results even from severely corrupted signals. Thus, this semantic coding pipeline ensures resilient data transmission in end-to-end machine-centric semantic communication systems.

\section{Open Issues and Future Directions}

\subsection{Semantic Coding Mitigates Semantic Uncertainty}
Analyzing semantic uncertainty in intent-driven communication systems has been challenging due to diverse requirements and varying metrics across  tasks and services. While there are a thousand Hamlets in a thousand people's eyes, semantic understanding in technical aspect can be simplified by decomposing semantic encoding and decoding (interpretation) processes: ignoring individual preferences during encoding through general semantic representation, \ie let data tell humans what it objectively is, then recovering useful content based on specific requirements at receiver, \ie make humans dominate how data is subjectively employed. This decoupled architecture, founded on semantic coding, mitigates ambiguous semantic uncertainty and establishes unified data-driven coding schemes as the foundation for task-oriented transmission.

\subsection{Semantic Coding Empowers Resilient Communication}
Achieving efficient and reliable communication in extremely poor and dynamic wireless channel conditions (\eg deep space, underwater, satellite, and high-mobility communications) remains both valuable and challenging. Simple dimensionality reduction or signal-level compression of data sources proves insufficient for real-time and effective transmission\cite{dai-deep}, lacking truly meaningful information extraction based on semantic understanding. Semantic coding is assumed to enable compact semantic representation while ensuring effective task-relevant content recovery by using the strong prior information modeled in the semantic latent space. Moreover, it facilitates resilient transmission of compact representations, effortlessly protecting significant information regardless of signal-level distortions, and can even self-heal corrupted transmitted signals via the generative prior-aided reconstruction, catering to the enduring pursuit of extreme scenario communications.

\subsection{Semantic Coding Catalyzes Generative Communication}
Generative communication is emerging as a new paradigm aimed at regenerating task-oriented content semantically equivalent to transmitted data through generative models \cite{ren-kb}, thereby reducing data traffic and offering versatility for novel applications. Semantic coding inherently operates from this generative perspective, extracting general semantic representations for generator-based content recovery while incorporating additional conditions from knowledge bases or prompts according to specific task requirements. Meanwhile, this coding scheme facilitates generative semantic communication by providing a unified technical pipeline for token-based semantic understanding and transmission. Compact semantic tokens can serve as a catalyst in balancing the reconstruction-generation trade-off \cite{yao-rec&gen}. Moreover, semantic coding procedure, encompassing tokenization, reorganization, and optional quantization, is well-suited for downstream tasks involving multiple modalities or embodied AI agents, and poised to inspire future explorations in generative semantic communication.

\section{Conclusion}
This article introduces semantic coding to address the intrinsic limitations of general neural coding, which serves as a fundamental technical guideline for constructing the coding workflow in end-to-end semantic communication systems that convey not merely deep features but the inherent semantic meaning of raw data.

Semantic coding includes three crucial operations on data sources: tokenization, reorganization, and optional quantization. Through examining the theoretical framework, technical pipeline, and practical applications of semantic coding, we establish that semantic communication extends beyond feature transmission to the conveyance of compact semantic representations via context-aware coding schemes. In essence, this approach effectively decomposes semantic understanding and general representation at transmitter from intended semantic interpretation at receiver, embodying the unity of commonality and individuality in semantic communication.

\bibliographystyle{IEEEbib}
\bibliography{ref}

\end{document}